\documentclass[preprint, 12pt, a4paper]{elsarticle}
\usepackage{amsmath,amssymb,amsfonts}
\usepackage{algorithmic}
\usepackage{graphicx}
\usepackage{ulem}
\usepackage[linesnumbered,lined,boxed,commentsnumbered,ruled]{algorithm2e}
\def\BibTeX{{\rm B\kern-.05em{\sc i\kern-.025em b}\kern-.08em
		T\kern-.1667em\lower.7ex\hbox{E}\kern-.125emX}}
\newcounter{bla}

\usepackage{listings}
\usepackage{xcolor}
\usepackage{textcomp}

\usepackage{caption}

\usepackage{natbib}

\definecolor{green}{rgb}{0,0.5,0}
\definecolor{green_2}{rgb}{0.133,0.545,0.133}
\lstnewenvironment{python}[1][]{
	\lstset{
		language=python,
		basicstyle=\tiny\sffamily,
		backgroundcolor=\color{gray!10},
		stringstyle=\color{red},
		showstringspaces=false,
		alsoletter={1234567890},
		otherkeywords={\ , \}, \{},
		keywordstyle=\color{blue},
		emph={access,and,break,class,continue,def,del,elif ,else,%
			except,exec,finally,for,from,global,if,import,in,i s,%
			lambda,not,or,pass,print,raise,return,try,while},
		emphstyle=\color{black}\bfseries,
		emph={[2]True, False, None, self},
		emphstyle=[2]\color{green},
		emph={[3]from, import, as},
		emphstyle=[3]\color{blue},
		upquote=true,
		morecomment=[s]{"""}{"""},
		commentstyle=\color{green_2}\slshape,
		emph={[4]1, 2, 3, 4, 5, 6, 7, 8, 9, 0},
		emphstyle=[4]\color{blue},
		literate=*{:}{{\textcolor{blue}:}}{1}%
		{=}{{\textcolor{blue}=}}{1}%
		{-}{{\textcolor{blue}-}}{1}%
		{+}{{\textcolor{blue}+}}{1}%
		{*}{{\textcolor{blue}*}}{1}%
		{!}{{\textcolor{blue}!}}{1}%
		{(}{{\textcolor{blue}(}}{1}%
		{)}{{\textcolor{blue})}}{1}%
		{[}{{\textcolor{blue}[}}{1}%
		{]}{{\textcolor{blue}]}}{1}%
		{<}{{\textcolor{blue}<}}{1}%
		{>}{{\textcolor{blue}>}}{1},%
		framexleftmargin=1mm, 
		framextopmargin=1mm, 
		frame=tb, 
		rulesepcolor=\color{gray},#1
}}{}

\begin{document}
\begin{frontmatter}



\title{Manapy: MPI-Based framework for solving partial differential equations using finite-volume on unstructured-grid}


\author[a]{I. Kissami\corref{author}}
\author[a]{A.Ratnani}

\cortext[author] {Corresponding author.\\\textit{E-mail address:} imad.kissami@um6p.ma}
\address[a]{MSDA, Mohammed VI Polytechnic University  Lot 660, 43150 Ben Guerir, Maroc}

\begin{abstract}
Manapy is a parallel, unstructured, finite-volume based solver for the solution of partial differential equations (PDE). The framework is written using Python, it is object-oriented, and is organized in such a way that it is easy to understand and modify. In this paper, we present the parallel implementation and scalability of the differential operators used on a general case of PDE. The performance of massively parallel direct and iterative methods for solving large sparse systems of linear equations in plasma physics is evaluated on a latest high performance computing system, and 3D test cases for plasma physics are presented.	
\end{abstract}

\begin{keyword}
\uline{MPI\sep Finite Volume Method\sep CFD\sep MUMPS\sep PETSc\sep Plasma physics.}
\end{keyword}

\end{frontmatter}

\section{Introduction}

Computational Fluid Dynamics(CFD) deals with the numerical solution of the governing equations of fluid dynamics. This discipline is widely used in different applications, such as environmental industries, aeronautics, plasma physics and automotive. In addition to that, it's an essential tool for academic research in any field that deals with fluid dynamics, parallel computing, and numerical methods. 




With the aim at offering an easy to understand and easy to modify CFD solver, and at the same time with discretization capabilities similar to those available in the most popular solvers available in the web (e.g., OpenFOAM \cite{chen2014openfoam}, SOLIDWORKS \cite{solidworks2005solidworks} and Ansys Fluent \cite{fluent2015ansys}), a wide range of modules and APIs implementing new efficient simulation methods on current high-performance computing(HPC) system are performed using different classes of numerical methods.

For finite-element based frameworks, we can find "Multiphysics Object-Oriented Simulation Environment" (MOOSE) \cite{icenhour2018multi, permann2020moose}, written in C++ and which scales properly on up to 32,768 MPI cores using hexahedral elements. The FEniCS framework \cite{alnaes2015fenics, richardson_wells_2015} contains high-level Python and C++ interfaces and includes several features for the automated, efficient solution of differential equations which are solved in parallel using MPI. Cimrman et al. \cite{CimrmanLukesRohan} presented SfePy (simple finite elements in Python) a software dedicated to solving a wide range of problems described by partial differential equations, that is, mainly, written in Python. In their work, they focused on a subpackage intended for complex multiscale numerical simulations. This software was successfully employed for various problems in biomechanics and materials science based on the theory of homogenization, which is suitable for multiphysical and multiscale simulations \cite{Lukes2020HomogenizationOL, Rohan2021HomogenizationOT}.

An other class of numerical method, Lattice Boltzmann, was used on the massively parallel "widely applicable Lattice Boltzmann from Erlangen" (waLBerla) framework \cite{BARTUSCHAT2018147} designed to efficiently run different stencil based codes on current HPC systems. To parallelize these calculations, waLBerla uses block structured grids and shows in \cite{BAUER2021478} good scaling up to 262,144 MPI cores.

Finally, a wide range of frameworks are developed using the finite-volume method (FVM), used to solve the conservation laws, which are the base of the governing equations of fluid dynamic. Guyer et al. \cite{Fipy} have been developing an interesting PDE solver, called FiPy, written in Python. Their framework aim to help improving performance, especially for large and complex problems, using parallel computing and efficient matrix preconditioners and solvers. In \cite{ALINOVI2021100655}, FLUBIO, a Fortran based, an unstructured, parallel, finite-volume based Navier–Stokes and convection–diffusion like equations solver for teaching and research purposes is presented. The use of these two frameworks remains very relevant but no performance study has been done. 

With the same aim, we introduce Manapy \footnote{https://github.com/pyccel/manapy}, Python3 FV framework with high temporal and spacial discretizations, dealing with both 2D and 3D unstructured-grid.  All functions are accelerated using either Numba \cite{10.1145/2833157.2833162} or Pyccel  \footnote{https://github.com/pyccel/pyccel}(generates both C and Fortran functions). Manapy is based on MPI parallelism and use METIS/PARMETIS \cite{Karypis:1998:FHQ:305219.305248} for mesh decomposition. The framework give choice to use either direct method using pymumps \footnote{https://github.com/pymumps/pymumps} (with adding functions to deal with distributed matrix and rhs) or iterative one using petsc4py (with different type of preconditioner) \cite{DALCIN20111124, osti_1483828} for solving Poisson equation, Matplotlib \cite{4160265} for 2D plots, Paraview (vtu, h5, pvtu) \cite{AHRENS2005717} for 2D/3D plots. Different models have been developed  using Manapy with Master and PhD. students (2D/3D Poisson equation, 2D Shallow Water \cite{10.1007/978-3-030-43651-3_42, 10.1007/978-3-030-43651-3_42}, 2D Shallow Water Magnetohydrodynamics \cite{atmos11040314}, 2D/3D Streamer discharge \cite{BENKHALDOUN20124623, Fort2019}, 2D Transient, incompressible, Navier–Stokes solver using the PISO algorithm \cite{ISSA198640}).

In this paper, the technical aspects, numerical discretization, optimization strategies and the performance of the  "MPI-Based framework for solving PDEs using FVM on unstructured-grid" (Manapy) framework are described. Afterwards, two 3D examples for streamer discharge are presented, the performance and scalability of the code are shown and discussed. Example of solving 3D Poisson's equation is given in the annexes.

%

\section{Mathematical Model}\label{sec2}
As a starting point, we consider both convective-diffusive equation \eqref{eq:1} and Poisson's equation \eqref{eq:1.1}, written as:
\begin{equation}
\begin{aligned}
\frac{\partial u}{\partial t}+\nabla\cdot(u\vec{V}) -  \nabla \cdot (D\vec{\nabla} u) = S\\
\end{aligned}
\label{eq:1}
\end{equation}
\begin{equation}
\begin{aligned}
\Delta P=f, \\ \hspace{.5cm}
\end{aligned}
\label{eq:1.1}
\end{equation}
\subsection{Temporal and Spacial discretization of the equation \eqref{eq:1}}
\label{sec3}
We use a FV approach (equation \eqref{eq:2}), in which the following quantities are defined at cell centers: the solution $u$, the velocity $V$. The fluxes are defined at cell faces. 
\begin{equation}
\begin{aligned}
\iint_{T_i} \frac{du}{dt} \; dV  + \iint_{T_i} \nabla \cdot(u\vec{V})\; dV - \iint_{T_i} \nabla \cdot (D\vec{\nabla} u) \; dV = \iint_{T_i} S \; dV
\label{eq:2}
\end{aligned}
\end{equation}
By using Green's formula and dividing by the volume, the equation \eqref{eq:2} leads to
\begin{equation}
\begin{aligned}
\frac{du}{dt} + \frac{1}{\mu_i} \oint_{ \partial T_i} u\vec{V}\cdot\vec{n} \; ds - \frac{1}{\mu_i} \oint_{\partial T_i} D \vec{\nabla} u\cdot \vec{n} \; ds = S
\label{eq:3}
\end{aligned}
\end{equation}
\noindent with the unit normal vector $\vec{n}$ and $\mu_i$ the volume of cell $T_i$. Now we approximate the curvilinear integral by a summation. So one obtains for a cell $T_i$.
\begin{equation}
\begin{aligned}
\frac{du}{dt}= 
-\frac{1}{\mu_i}\underbrace{\sum_{j=1}^{m}{u}_{ij}^n\vec{V}_{ij}\vec{n}_{ij}|\sigma_{ij}|}_{Rez\_conv} + \frac{1}{\mu_i}\underbrace{\sum_{j=1}^{m} D{\vec \nabla}{u}_{ij}^n\vec{n}_{ij}|\sigma_{ij}|}_{Rez\_dissip} + S_i^n \\
\label{eq:4}
\end{aligned}
\end{equation}
\noindent where m is the faces number of cell $T_i$, $\vec{n}_{ij}$ is the unit normal vector on the face $\sigma_{ij}$ (face between cells {$T_i$ and $T_j$) and $|\sigma_{ij}|$ is the face's surface.  Other variables denoted by subscript $ij$ represent variables on the face $\sigma_{ij}$. 
\paragraph{Convective flux discretization}
\noindent The convective flux ${u}_{ij}$ in the equation \eqref{eq:2} is computed using a given FV scheme. Here we consider the simple upwind for simplification, extended by a Van Leer’s type MUSCL algorithm along with Barth-Jespersen limiter in order to achieve a second order accuracy in space.
\begin{equation}
u_{ij}=
\left\{
\begin{aligned}
	{u}_i + \psi_i \cdot ({\nabla} {u}_i \cdot \vec{r_i})  \hspace{1.5cm}
	if (\vec{{V}_{ij}}\;.\;\vec{n_{ij}}) \geq 0\; \\
	{u}_j + \psi_j \cdot ({\nabla} {u}_j \cdot \vec{r_j})\hspace{2.5cm} otherwise
	\label{15}
\end{aligned}
\right.
\end{equation}
\vspace{.2cm}
\noindent where $\nabla{u}_i$, $\nabla{u}_j$ are $u$ gradients on cells $T_i$, $T_j$. These gradients are computed assuming that $u$ is a piecewise linear function and its value ${u}_i$ is in the center of gravity of the cell $T_i$. This linear function is computed by the least square method involving all neighboring cells of the vertices of $T_i$ (see algorithm \ref{algo:grad}). $r_i$ ($r_j$) is a vector coming from $T_i$($T_j$) center of gravity to face $\sigma_{ij}$ midpoint. $\psi_i$($\psi_j$) is the Barth-Jespersen limiter function. The parallel implementation of $\nabla{u}_i$ and $\nabla{u}_j$ is detailed in the algorithm \ref{algo:grad}, and figure \ref{fig:mesh_decomp} (projected in 2D), illustrates the whole information needed. E.g, for subdomain 1, the cell $cB1$ needs around node $B$; the ghost cell $gB1$, the halo cells $hB1$ and $hB2$ and the halo-ghost cell $hgB1$ coming from the subdomain 2. 
\IncMargin{1em}
\begin{algorithm}[h!]
\begin{footnotesize}
	C : number of cells;\\
	$A_{x}$, \hspace{.2cm}  $A_{y}$, \hspace{.2cm} $A_{z}$ and D: computed using mesh;
	\vspace{.2cm}
	
	\For{i:=1 \emph{\KwTo} C}{
		\For{j $\in$  inner cells  neighbor by node}{
			$J_x$ += ($A_x$ * ($u(j)$ - $u(i)$));\hspace{.2cm}
			$J_y$ += ($A_y$ * ($u(j)$ - $u(i)$));\hspace{.2cm}\\
			$J_z$ += ($A_z$ * ($u(j)$ - $u(i)$));
		}
		\For{h $\in$ halo cells neighbor by node}{
			$J_x$ += ($A_x$ * ($u(h)$ - $u(i)$));\hspace{.2cm}
			$J_y$ += ($A_y$ * ($u(h)$ - $u(i)$));\hspace{.2cm}\\
			$J_z$ += ($A_z$ * ($u(h)$ - $u(i)$));
		}
		\For{g $\in$ ghost cells neighbor by node}{
			$J_x$ += ($A_x$ * ($u(g)$ - $u(i)$));\hspace{.2cm}
			$J_y$ += ($A_y$ * ($u(g)$ - $u(i)$));\hspace{.2cm}\\
			$J_z$ += ($A_z$ * ($u(g)$ - $u(i)$));
		}
		\For{hg $\in$ haloghost cells neighbor by node}{
			$J_x$ += ($A_x$ * ($u(hg)$ - $u(i)$));\hspace{.2cm}
			$J_y$ += ($A_y$ * ($u(hg)$ - $u(i)$));\hspace{.2cm}\\
			$J_z$ += ($A_z$ * ($u(hg)$ - $u(i)$));
		}
		$(u)_x(i)$ =  $f(A_x, A_y, A_z, D)$; \hspace{.2cm}
		$(u)_y(i)$ =  $f(A_x, A_y, A_z, D)$;\\
		$(u)_z(i)$ =  $f(A_x, A_y, A_z, D)$;	
	}
	\caption{\footnotesize{Compute the gradient $\nabla u$ \big($(u)_x$, $(u)_y$, $(u)_z$\big) on cell $T_i$}}
	\label{algo:grad}
\end{footnotesize}
\end{algorithm}
\begin{figure}[h!]
	\caption{An example of a 3D (projected in 2D) mesh (left) and a mesh decomposition into four subdomains (right).}
	\centering
	\includegraphics[width=.7\textwidth]{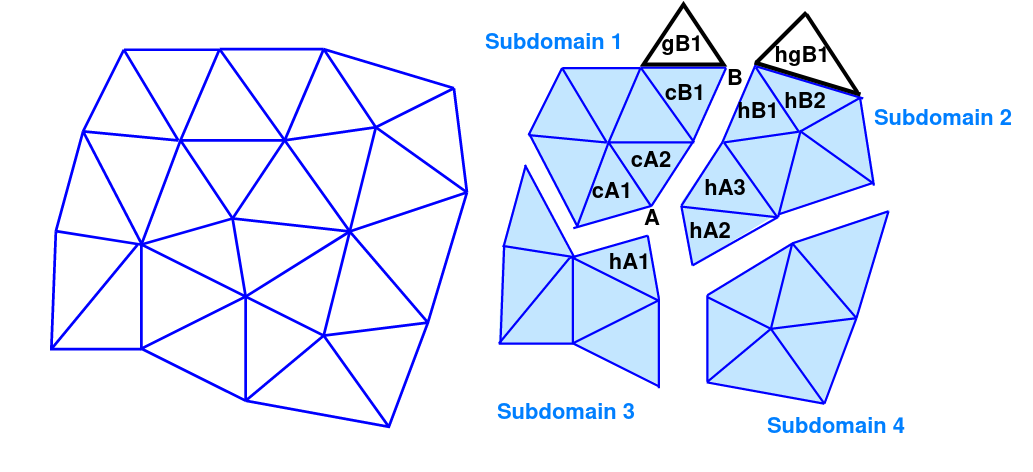}
	\label{fig:mesh_decomp}
\end{figure}

\paragraph{Diffusive flux discretization} The face gradient $\nabla {u}_{ij}$ in the dissipative part is approximated using the FV diamond scheme. To allow a simplification of $\nabla {u}_{ij}$ written in equation \eqref{eq:diamond}, we consider the general representation of tetrahedron illustrated in figure \ref{fig:diamond}. In our study, the tetrahedron base is triangle $(D=C)$.
\begin{equation}
\begin{split} 
	{\vec{\nabla}}{u}_{ij} = \frac{1}{3\mu(D_{\sigma_{ij}})}\bigg[
	\big(u(A) - u(C)\big) \vec{n}_{BRDL}|\sigma_{BRDL} | + \big(u(B) - \\
	u(D)\big)\vec{n}_{ALCR} |\sigma_{ALCR} | + \big(u(R) - u(L)\big)\vec{n}_{ij} |\sigma_{ij}|\bigg]
	\label{eq:diamond}
\end{split} 
\end{equation}
\begin{figure}[h!]
\caption{General representation of Diamond cell in 3D.}
\centering
\includegraphics[width=.35\textwidth]{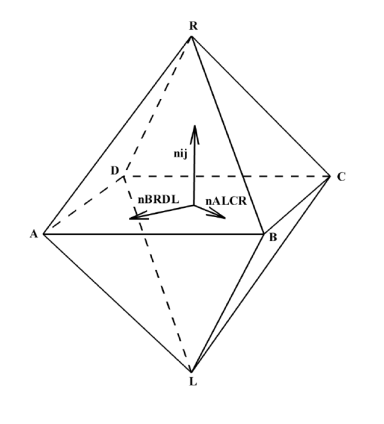}
\label{fig:diamond}
\end{figure}
\noindent The values $u(A)$, $u(B)$, $u(C)$ and $u(D)$ are computed by the least square method, detailed in the algorithm \ref{algo:ls}.
\IncMargin{1em}
\begin{algorithm}[!h]
\begin{footnotesize}
	$u_{A}$: double\;
	$N$: Number of nodes\;
	$Alpha$: weight coming from the least square method\;
	\vspace{.2cm}
	\For{n:=1 \emph{\KwTo} N }{
		\For{c $\in$  inner cells around node n}{
			$u_{A}(n)+=Alpha(c) * u(c)$\;
		}
		\For{h $\in$ halo cells around node}{
			$u_{A}(n)+=Alpha(h) * u(h)$\;
		}
		\For{g $\in$ ghost cells around node}{
			$u_{A}(n)+=Alpha(g) * u(g)$\;
		}
		\For{hg $\in$ haloghost cells around node}{
			$u_{A}(n)+=Alpha(hg) * u(hg)$\;
		}
	}
	\Return $u_{node}$\;
	\caption{\footnotesize{Least square interpolation (value on node A)}}
	\label{algo:ls}
\end{footnotesize}
\end{algorithm}
\paragraph{Temporal discretization}
\label{sec4}
\noindent Time stepping is performed using the three steps Runge-Kutta method.
\begin{eqnarray}
u^{(0)} = {u}_i^{(n)} \nonumber\\
{u}_i^{(k)} = {u}_i^{(n)} + \alpha_{(k)} Res\left({u}_i^{(k-1)}\right) \nonumber\\
{u}_i^{(n+1)} = {u}_i^{(n)}\nonumber\\
\label{eq:6}
\end{eqnarray}
\noindent with $k = 1,\;2,\;3$ and the coefficients $\alpha_{(1)} = 0.5$, $\alpha_{(2)} = 0.5$, $\alpha_{(3)} = 1$. The residual in	equation \eqref{eq:4} is the summation of convective, dissipative and source term fluxes.	
\subsection{Discretization of Poisson's equation \eqref{eq:1.1}} \label{sec5}
One of the key computational challenges in multiphysics simulations is quickly solving Poisson's equation $\nabla^2 P= f$, with high spatial resolution. In Manapy, the Poisson's equation
is discretized by a central type approximation which leads to a system of linear equations
\begin{equation}
\begin{aligned}
	\textbf{A}.\vec{P}^{n}=\vec{b}^{n}\label{14}
\end{aligned}
\end{equation}
A is a matrix of coefficients, $\vec{P}$ is a vector of unknowns (its dimension is equaled to the total number of cells) and $\vec{b}$ is a vector of right hand side. A row i in the matrix A corresponds to the cell $T_i$ . We use a similar FVM approximation as for diffusive terms in the equation \eqref{eq:4}:
\begin{equation}
\nabla^2 P = \frac{1}{\mu_i}\sum_{j=1}^{m} {\vec{\nabla}}{P}_{ij}\;\vec{n}_{ij}|\sigma_{ij}|
\label{eq:7}
\end{equation}
\noindent An approximation of the gradient ${\vec{\nabla}}{P}_{ij}$ is performed according to equation \eqref{eq:diamond}. 
\vspace{.2cm}
\begin{equation}
\begin{split}
	\nabla^2 P = \frac{1}{3\mu_i\mu(D_{\sigma_{ij}})}\bigg[
	\vec{n}_{BRDL}|\sigma_{BRDL} |\cdot\vec{n}_{ij}|\sigma_{ij}|\cdot P_A - \\
	\vec{n}_{BRDL}|\sigma_{BRDL} |\cdot\vec{n}_{ij}|\sigma_{ij}|\cdot P_C + \\	
	\vec{n}_{ALCR}|\sigma_{ALCR} |\cdot\vec{n}_{ij}|\sigma_{ij}|\cdot P_B - \\
	\vec{n}_{ALCR}|\sigma_{ALCR} |\cdot\vec{n}_{ij}|\sigma_{ij}|\cdot P_D + \\
	\vec{n}_{ij}|\sigma_{ij} |\cdot\vec{n}_{ij}|\sigma_{ij}|\cdot P_R - \\
	\vec{n}_{ij}|\sigma_{ij} |\cdot\vec{n}_{ij}|\sigma_{ij}|\cdot P_L \bigg]
	\label{eq:laplace}
\end{split}
\end{equation}

\noindent The values $P_A$ , $P_B$ , $P_C$ and $P_D$ come from the least square method.

\clearpage
\section{Manapy performance}

In this section we aim to validate the parallel implementations of the different finite volume operators (cell gradient, face gradient and least square interpolation) used for the convection-diffusion equation \eqref{eq:1}, as well as the resolution of the Poisson equation \eqref{eq:1.1} with both direct and iterative solvers. Table \ref{table:grids} shows the different grids used in this study.

\begin{table}[hbtp]
	\footnotesize
	\renewcommand{\arraystretch}{1}
	\setlength\tabcolsep{8pt}
	\centering
	\begin{center}
		\caption{Test grid characteristics.}\label{table:grids}
		\begin{tabular}{ll}
			\hline
			\noalign{\smallskip}
			Grid  & Grid size \\
			\hline 
			\noalign{\smallskip}
			G1  &  1,055,603 \\
			G15 &  15,462,236  \\
			G30 &  29,860,926  \\
		\end{tabular}
	\end{center}
\end{table}
\vspace{-.2cm}
\subsection{Working environment}
To realize all the experiments, we worked on the TOUBKAL cluster \footnote{https://ascc.um6p.ma/}  which is located at Mohammed VI Polytechnic University (Benguerir, Morocco). This machine contains about 1219 nodes with 178 GB RAM, each node have 2 sockets of CPU Intel Xeon Platinum 8276 Processor (38.5M Cache, 2.20 GHz and 28 Cores), interconnected by HDR Infiniband

\subsection{Halo data} The halo information sent each iteration are unknowns which depend on the given equation (for equation \eqref{eq:1}, unknowns are; solution $u$, velocity $\vec{V}(V_x, V_y, V_z)$ and cell gradient $\nabla u({u}_x, {u}_y, {u}_z)$. Table \ref{tab:metis_statistic} shows the halo cells and neighbors needed to elaborate the different FV operators in equation \eqref{eq:4} for the partition with the maximum cells account. 

\begin{table}[h!]
	\footnotesize
	\renewcommand{\arraystretch}{1}
	\setlength\tabcolsep{6pt}
	\begin{center}
		\caption{Statistics of the partition with maximum number of cells and neighboring depending on MPI cores number.}\label{tab:metis_statistic}
		\begin{tabular}{llllllllll}
			\noalign{\smallskip}
			\hline
			\noalign{\smallskip}
			Grid & \multicolumn{3}{c} {2048 MPI cores} & \multicolumn{3}{c} {8192 MPI cores} & \multicolumn{3}{c} {32768 MPI cores} \\
			& Inner & Halo & Neigh. & Inner & Halo & Neigh. & Inner & Halo & Neigh. \\
			\hline \\
			G1 & 500 & 1,469 & 20 & 125 & 1211 & 22 & 31&  988& 80\\
			G15 & 7,776 & 6,969 & 18 & 1,832 & 3,049 &18 & 458 & 1,867 & 26   \\
			G30 & 14,011 & 9,455 & 22 & 3,503 & 4,096& 19 &  825 & 2,218 & 23\\
			\hline
		\end{tabular}
	\end{center}
\end{table}

\subsection{Strong scaling for the FV operators implementation}
Tables \ref{table:parts1M}, \ref{table:parts15M}, \ref{table:parts30M} show computational time (s) for the FV operators using G1, G15, G30 respectively for one iteration. Figures \ref{fig:speedup1M},
\ref{fig:speedup15M} and \ref{fig:speedup30M} represent the strong speedup for the FV operators using G1, G15, G30 respectively. The good scaling depends on two major reasons; 
\begin{itemize}
	\item cache optimization; which depends on the grid size and MPI cores, i.e higher MPI cores leads to small partitions means that more information fit on the cache memory (cache hit).
	\item computational optimization; more the number of cores increase more the computational cost becomes small. However, for a very small number of cells per core we clearly see additional overhead of the code, mainly due to the large communication surface (halo cells).
\end{itemize}

Unfortunately using G1 the scaling is limited to 4096 MPI cores, because the cell's number per core is very small (no cache optimization, $500$, $248$ and $123$ using 2048, 4098, 8192  MPI cores respectively), and more communications ($1524$, $1469$ and $1211$, using 2048, 4098, 8192  MPI cores respectively). 

However using G30, the strong speedup can be 3 times higher than ideal one (example of cell gradient computation), because we take advantage for both optimizations.

\begin{table}[htbp]
	\footnotesize
	\begin{center}
		\caption{Computational time (s) for the FV operators using grid G1.}\label{table:parts1M}
		\begin{tabular}{llllll}
			\hline
			\noalign{\smallskip}
			Cores & Nodes & Cell Gradient & Face Gradient & LS interpolation  \\
			\hline
			\noalign{\smallskip}
			1 & 1 & 1.18212 & 0.09743 & 0.05960 \\
			2 & 1 & 0.56895 & 0.05151 & 0.03058 \\
			4 & 1 & 0.29468 & 0.02811 & 0.01639 \\
			8 & 1 & 0.16104 & 0.01282 & 0.00917 \\
			16 & 1 & 0.07927 & 0.00626 & 0.00490 \\
			32 & 1 & 0.04124 & 0.00488 & 0.00280 \\
			64 & 2 & 0.01687 & 0.00201 & 0.00132 \\
			128 & 3 & 0.00796 & 0.00069 & 0.00057 \\
			256 & 5 & 0.00358 & 0.00043 & 0.00021 \\
			512 & 10 & 0.00172 & 0.00025 & 0.00010 \\
			1024 & 19 & 0.00086 & 0.00012 & 0.00005 \\
			2048 & 37 & 0.00044 & 0.00008 & 0.00003 \\
			4096 & 74 & 0.00025 & 0.00006 & 0.00002 \\
			8192 & 147 & 0.00022 & 0.00004 & 0.00002 \\
		\end{tabular}
	\end{center}
\end{table}
\begin{figure}[htbp]
	\centering
	\caption{Strong Speedup for the FV operators using grid G1.}
	\includegraphics[width=.7\textwidth]{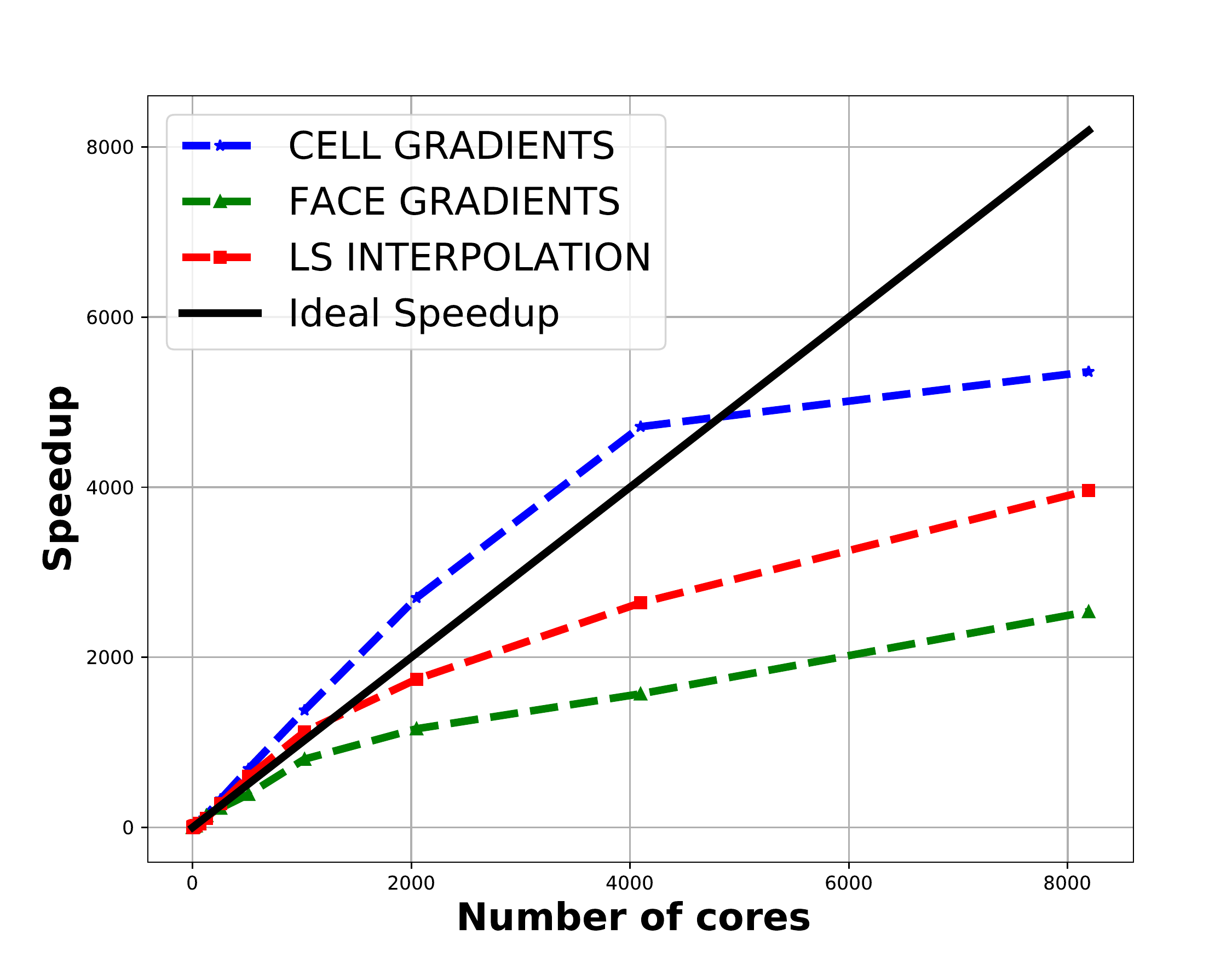}
	\label{fig:speedup1M}
\end{figure}

\begin{table}[htbp]
	\footnotesize
	\begin{center}
		\caption{Computational time (s) for the FV operators using grid G15.}\label{table:parts15M}
		\begin{tabular}{lllll}
			\hline
			\noalign{\smallskip}
			Cores & Cell Gradient & Face Gradient & LS interpolation  \\
			\hline
			\noalign{\smallskip}
			1 & 37.2615 & 1.92517 & 1.35792 \\
			2 & 18.3247 & 0.90603 & 0.65020 \\
			4 & 10.9948 & 0.53237 & 0.38328 \\
			8 & 6.03951 & 0.30884 & 0.21023 \\
			16 & 3.59826 & 0.17669 & 0.13934 \\
			32 & 2.12073 & 0.11027 & 0.08537 \\
			64 & 0.85504 & 0.05451 & 0.03491 \\
			128 & 0.35127 & 0.02529 & 0.01569 \\
			256 & 0.11843 & 0.00950 & 0.00642 \\
			512 & 0.04186 & 0.00427 & 0.00270 \\
			1024 & 0.01722 & 0.00176 & 0.00123 \\
			2048 & 0.00731 & 0.00075 & 0.00051 \\
			4096 & 0.00333 & 0.00049 & 0.00020 \\
			8192 & 0.00164 & 0.00031 & 0.00011 \\
			16384 & 0.00089 & 0.00017 & 0.00007 \\
			32768 & 0.00053 & 0.00019 & 0.00005 \\
		\end{tabular}
	\end{center}
\end{table}
\begin{figure}[htbp]
	\centering
	\caption{Strong Speedup for the FV operators using grid G15.}
	\includegraphics[width=.7\textwidth]{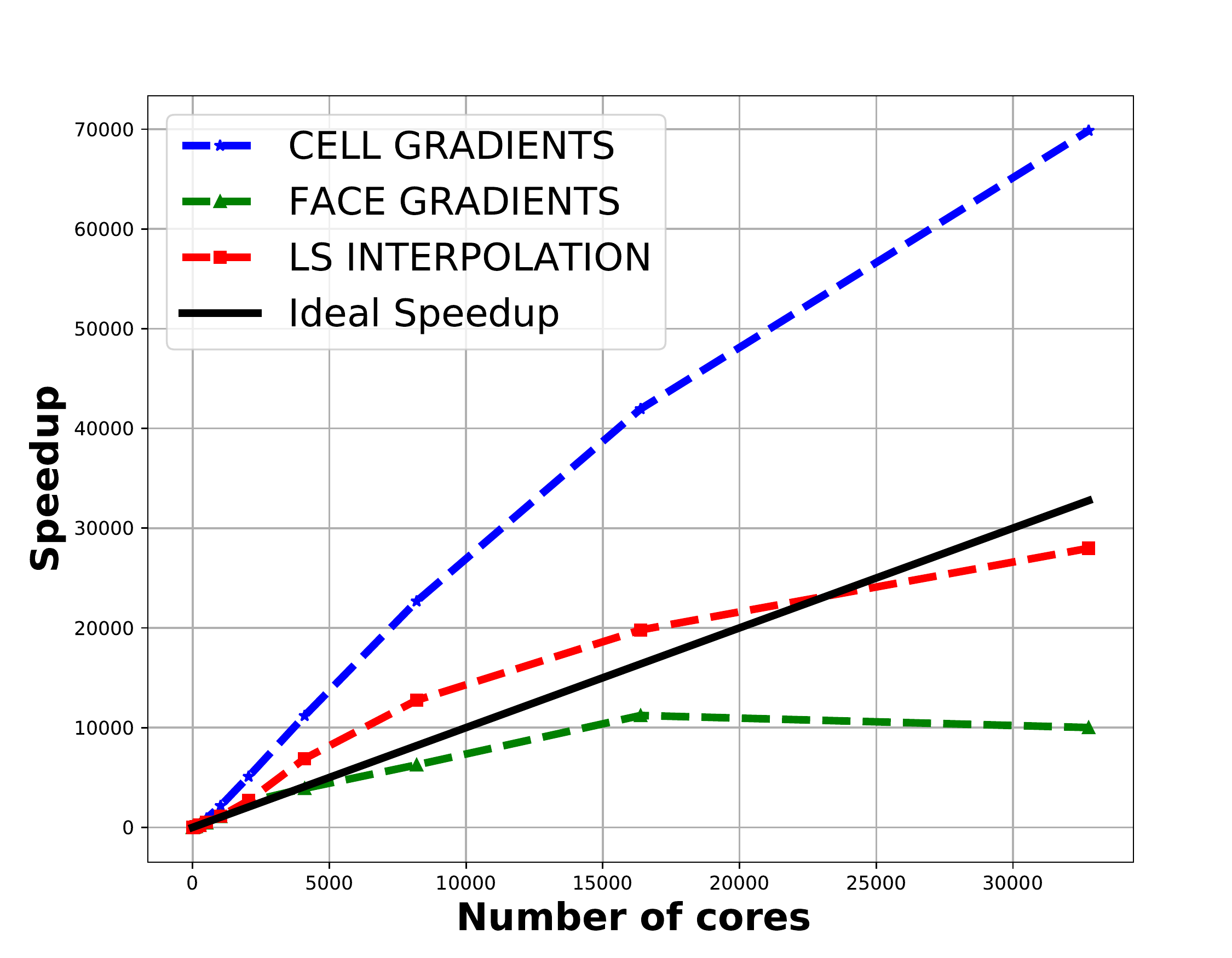}
	\label{fig:speedup15M}
\end{figure}

\begin{table}[htbp]
	\footnotesize
	\begin{center}
		\caption{Computational time (s) for the FV operators using grid G30.}\label{table:parts30M}
		\begin{tabular}{lllll}
			\hline
			\noalign{\smallskip}
			Cores  & Cell Gradient & Face Gradient & LS interpolation  \\
			\hline
			\noalign{\smallskip}
			1 & 80.9370 & 3.90363 & 2.81876 \\
			2 & 40.7047 & 2.43486 & 1.52669 \\
			4 & 24.2443 & 1.38506 & 0.84177 \\
			8 & 12.1788 & 0.78981 & 0.44978 \\
			16 & 6.92798 & 0.44197 & 0.25366 \\
			32 & 3.81057 & 0.28888 & 0.16145 \\
			64 & 1.74759 & 0.10506 & 0.07115 \\
			128 & 0.77727 & 0.04988 & 0.03198 \\
			256 & 0.30856 & 0.02000 & 0.01369 \\
			512 & 0.10685 & 0.00861 & 0.00592 \\
			1024 & 0.03725 & 0.00366 & 0.00241 \\
			2048 & 0.01482 & 0.00153 & 0.00111 \\
			4096 & 0.00654 & 0.00072 & 0.00043 \\
			8192 & 0.00303 & 0.00047 & 0.00018 \\
			16384 & 0.00152 & 0.00030 & 0.00011 \\
			32768 & 0.00085 & 0.00024 & 0.00007 \\
		\end{tabular}
	\end{center}
\end{table}

\newpage

\begin{figure}[htbp]
	\centering
	\caption{Strong Speedup for the FV operators using grid G30.}
	\includegraphics[width=.7\textwidth]{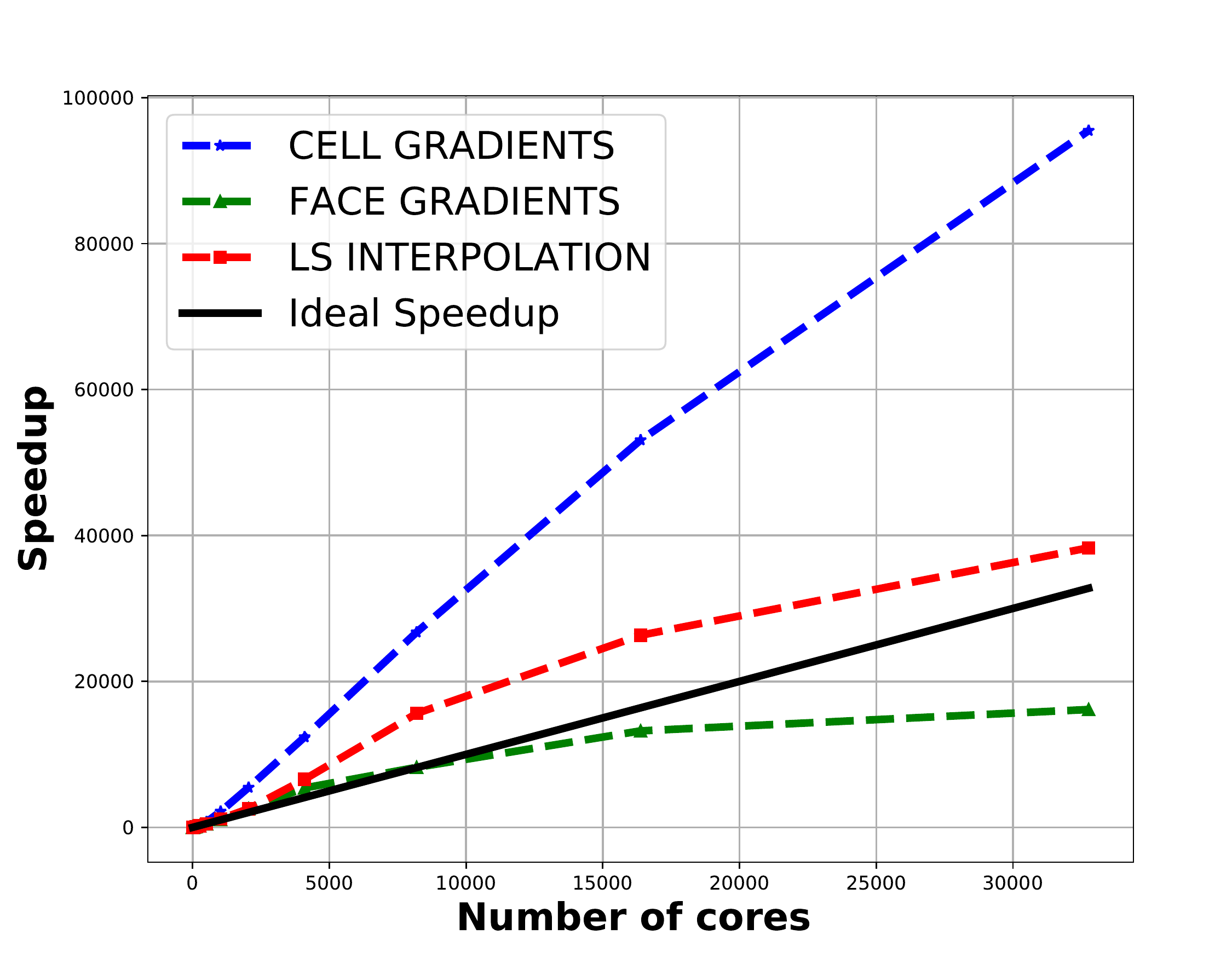}
	\label{fig:speedup30M}
\end{figure}

\subsection{Poisson's equation} 
\subsubsection{Parallel assembly for 3D Matrix} 
In Manapy, every processor compute its submatrix, the difficulty in such approach using the FV diamond scheme is dealing with the halo and haloghost cells around each node. Table \ref{table:matrices}, sum up the test matrices varying from 1 to 30 million degrees of freedoms (DOFs); more than 1 Billion nonzeros (64-bit double) for M15 and more than 2 Billion nonzeros for M30. Due to large memory requirements, M15 failed with less than 512 cores and M30 with less than 1024 cores. Table \ref{table:assembly} summarizes the computational times (s) for assembling the 3D matrices M1, M15 and M30. E.g. The speedup for assembling M1 using 16384 MPI cores is 20598$\gg$16384 (ideal speedup); this speedup is explained by both MPI and cache optimizations.

\begin{table}[h!]
	\footnotesize
	\renewcommand{\arraystretch}{1}
	\setlength\tabcolsep{8pt}
	\centering
	\begin{center}
		\caption{Test matrix characteristics.}\label{table:matrices}
		\begin{tabular}{lll}
			\hline
			\noalign{\smallskip}
			Matrix  & Grid size & Nonzeros \\
			\hline
			\noalign{\smallskip}
			M1  &  1,055,603 & 78,850,483 \\
			M15 &  15,462,236 & 1,176,894,832 \\
			M30 &  29,860,926 & 2,245,141,202 \\
		\end{tabular}
	\end{center}
\end{table}

\begin{table}[h!]
	\footnotesize
	\renewcommand{\arraystretch}{1}
	\setlength\tabcolsep{8pt}
	\begin{center}
		\caption{Computational time (s) for the Matrices assembly for 3D Poisson's equation.}\label{table:assembly}
		\begin{tabular}{llll}
			\hline
			\noalign{\smallskip}
			Cores  & M1 & M15 & M30  \\
			\hline
			\noalign{\smallskip}
			1     & 117.41 & F & F \\
			64    & 1.7792 & F & F \\
			128   & 0.8780 & F & F \\
			256   & 0.4221 & F & F \\
			512   & 0.2017 & 4.0127 & F \\
			1024  & 0.1030 & 1.9143 & 4.4896 \\
			2048  & 0.0461 & 0.9083 & 1.7574 \\
			4096  & 0.0211 & 0.4419 & 0.8202 \\
			8192  & 0.0103 & 0.1935& 0.3683 \\
			16384 & \bf{0.0057} & 0.0988 & 0.1812 \\ 
			32768  & 0.0066 & \bf{0.0504} & \bf{0.0881} \\ 
		\end{tabular}
	\end{center}
\end{table}

\subsubsection{Solving Poisson equation}
Both direct and iterative methods for solving linear system with sparse matrices are performed in Manapy (see implementation example in \ref{manapyexample}). Direct solvers generally perform better than iterative solvers on lower core counts, while iterative solvers are scaling better and approaching the performance of direct solvers on higher core count. However, as the size of tests increases, the memory requirement for direct solvers becomes an obstacle. Tables \ref{table:comp1} recapitulates the comparison between MUMPS and PETSC (FGMRES with different type of pre-conditioner) for test case M1. Figure \ref{fig:mumpsvspetsc}, shows that petsc using FGMRES with GAMG scales better for this kind of resolution. Due to 64-bit double overflow, MUMPS failed at M15, that's why we focus only on petsc using FGMRES with GAMG and PBJacobi. Table \ref{table:comp2} sum up the computational time (s). Figures \ref{fig:speeduppoisson1} and \ref{fig:speeduppoisson2} show that more matrix is large more petsc solver achieve a good scaling with higher core counts. 

\begin{table}[h!]
\footnotesize
\renewcommand{\arraystretch}{1}
\setlength\tabcolsep{6pt}
\begin{center}
	\caption{Solution times in seconds and iteration number for M1 matrix, with different solver and preconditioner combinations.}\label{table:comp1}
	\begin{tabular}{llllllll}
		\hline
		\noalign{\smallskip}
		Cores  & {MUMPS} & &\multicolumn{4}{l} {PETSc}  \\
		\cline{4-7}
		\noalign{\smallskip}
		&&   & GAMG(14) & Jacobi(607) & ASM(87) & Hypre(7)\\
		\hline 
		\noalign{\smallskip}
		1 & 12.87 && 34.67 & 194.1 & 26.52 & 31.77 \\
		2 & 7.159 && 16.67 & 102.2 & 13.95 & 16.72 \\
		4 & 4.577 && 9.970 & 56.78 & 7.754 & 9.286 \\
		8 & 3.022 && 4.527 & 25.88 & 6.266 & 5.477 \\
		16 & 2.594 && 1.809 & 12.14 & 4.098 & 2.795 \\
		32 & 2.188 && 1.426 & 9.851 & 4.297 & 1.988 \\
		64 & 1.228 && 0.726 & 5.518 & 3.559 & 1.081 \\
		128 & 1.047 && 0.516 & 4.016 & 3.291 & 0.790 \\
		256 & 1.045 && 0.398 & 2.844 & 3.899 & 0.636 \\
		\bf{512} & \bf{1.138} && \bf{0.351} & \bf{2.637} & \bf{3.301} & \bf{0.597} \\
		1024 & 1.311 && 0.506 & 3.470 & 1.646 & 0.988 \\
		\hline
	\end{tabular}
\end{center}
\end{table}
\begin{figure}[h!]
\caption{Speedup of solving 3D Poisson's equation for M1 with different solver and preconditioner combinations}
\centering
\includegraphics[width=.65\textwidth]{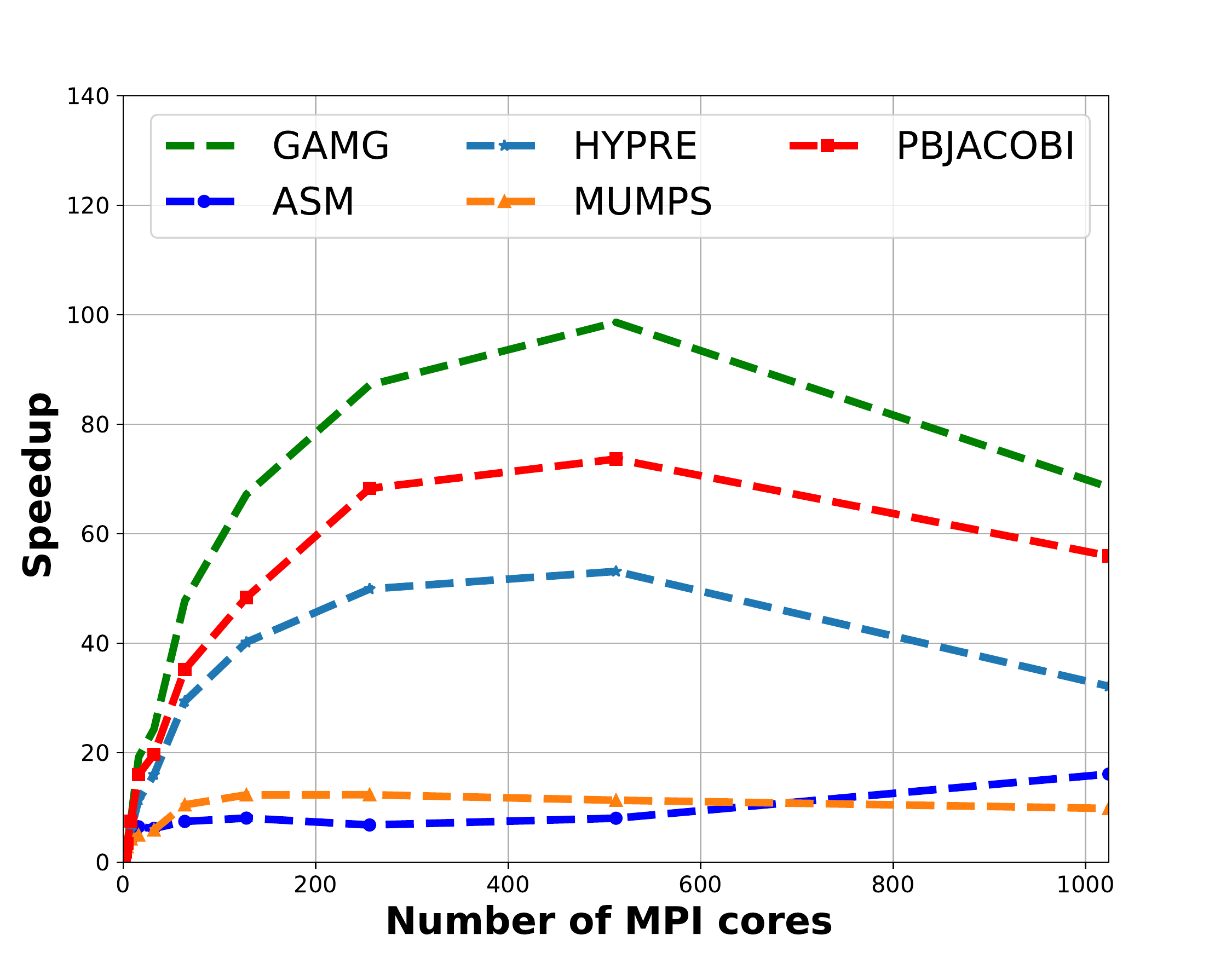}
\label{fig:mumpsvspetsc}
\end{figure}

\begin{table}[h!]
\footnotesize
\renewcommand{\arraystretch}{1}
\setlength\tabcolsep{4pt}
\begin{center}
	\caption{Solution times in seconds and iteration number for M15 and M30 matrices using FGMRES with tolerance = $10e^{-10}$, with different preconditioner.}\label{table:comp2}
	\begin{tabular}{llllllll}
		\hline
		\noalign{\smallskip}
		cores & & \multicolumn{2}{l} {M15} & &\multicolumn{2}{l} {M30}  \\
		\cline{1-3}
		\cline{4-7}
		\noalign{\smallskip}
		&& GAMG(17) & PBjacobi(1080)  & & GAMG(18) & PBjacobi(1081) \\
		\hline 
		\noalign{\smallskip}
		512 & & 80.96 & 3816   & &F & F\\
		1024 & & 40.51 & 1847  & &67.59 & 3410 \\
		2048 & & 24.14 & 984.9 & &32.18  & 1753\\
		4096 & & \bf{12.13} & \bf797.1 & &14.86  & 974.3\\
		8192 & & 25.22 & 1001 & & \bf{8.744}& \bf{608.9}\\
		\hline
	\end{tabular}
\end{center}
\end{table}
\begin{figure}[h!]
\centering
\includegraphics[width=.65\textwidth]{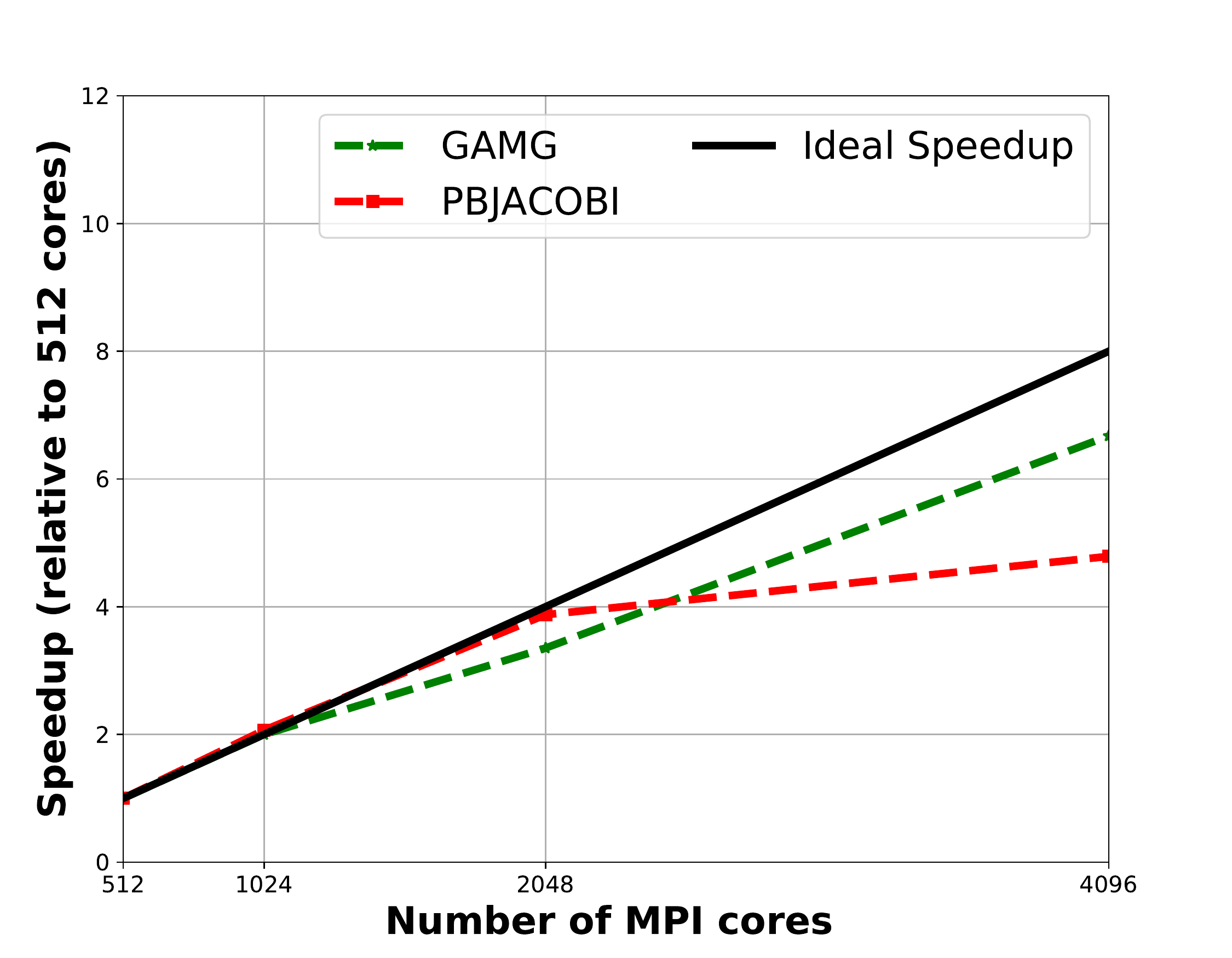}
\caption{Speedup of solving 3D Poisson's equation for M15 using PETSc with preconditioner combinations.}
\label{fig:speeduppoisson1}
\end{figure}
\begin{figure}[h!]
\centering
\includegraphics[width=.65\textwidth]{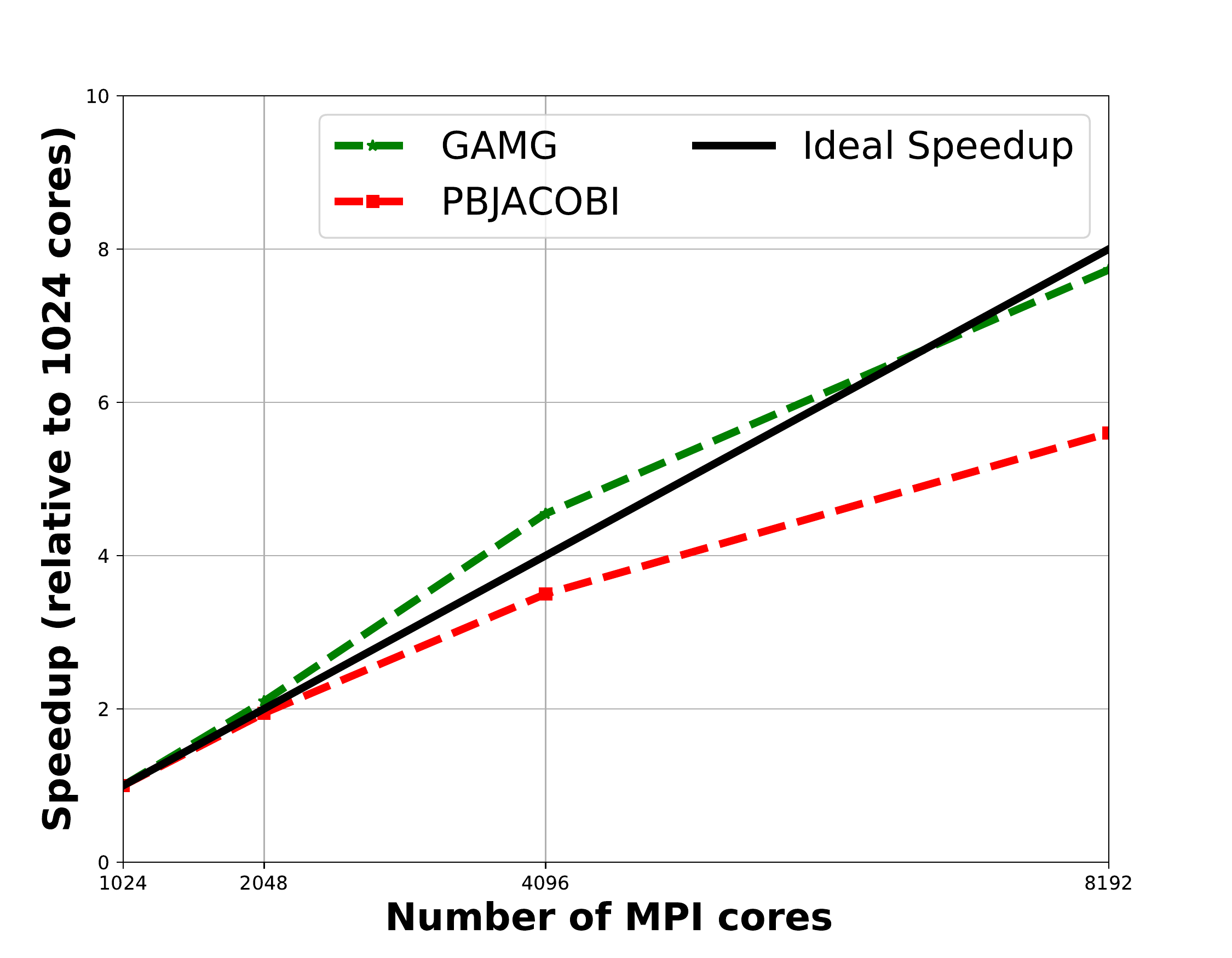}
\caption{Speedup of solving 3D Poisson's equation for M30  using PETSc with preconditioner combinations.}
\label{fig:speeduppoisson2}
\end{figure}

\section{Representing the 3D streamer model}\label{sec:tests}
To concretize these implementations, we choose to test the 3D Streamer model \cite{BENKHALDOUN20124623, Fort2019}, which couple both equations \eqref{eq:1} and \eqref{eq:1.1}. Streamers are the precursors of sparks, lightning leaders, sprites and they are also used in diverse applications in plasma technology, such as surface processing \cite{_ern_k_2011}, environmental applications \cite{Joshi2013}, catalysis \cite{NOZAKI201329}, sterilization and disinfection \cite{879360}. The challenge in this 3D simulation lies in the fine meshes needed to simulate rapid variations in the solution, because streamer discharges propagate at high speeds, e.g., at $10^6 m/s$. 

\subsection{Streamer discharge}
\noindent We consider the 3D problem of the discharge propagation in a homogeneous electric field described by equation \eqref{eq:10}. A cylinder domain (Grid G30) $\Omega = [L = 0.5 , r = 0.25] cm$ is considered. The convection-diffusion-reaction equation for the electron density is discretized using second order in space and time. The computation end time is $t=20.5$ ns, and total number of iterations is 25000. 
\begin{eqnarray}
\frac{\partial n_e}{\partial t} +
\nabla \cdot \left(n_e\vec{v_e}\right) - \nabla \cdot \left(D_e\nabla n_e\right) = S_e,\nonumber\\ \label{eq:10}
\ \ \frac{\partial n_p}{\partial t} = S_e,\\\nonumber
\nabla^2 V = -\frac{e}{\epsilon} \left(n_p - n_e\right)\ , 
\ \ \vec E = - \nabla V \ , \\ \nonumber 
\end{eqnarray}
\noindent $v_e$ is the electron velocity described in equation \eqref{eq:velo}, $\alpha$ is the effective ionization coefficient, $D_e$ the electron diffusion coefficient described in equation \eqref{eq:diff} and $S_e = \alpha \cdot ||\vec{v_e}|| \cdot n_e$.
The fluid equations are coupled to the electrostatic field, $V$ and $\vec{E}$ denote the electric potential and electric field, respectively, $\epsilon$ the permittivity of vacuum and e the elementary charge.
\begin{equation}
v_e = -\left[ C_1(\vec{E})\cdot\frac{||\vec{E}||}{N} + C_2(\vec{E})\right]
\cdot\frac{\vec{E}}{||\vec{E}||}
\label{eq:velo}
\end{equation}
\begin{equation}
D_e = -\left[ 0.3341\cdot10^9\cdot\frac{||\vec{E}||}{N}^{0.54069} 
\right]
\cdot\frac{\vec{v_e}}{||\vec{E}||}
\label{eq:diff}
\end{equation}
\noindent Where the ratio 
\begin{Large}
$\frac{\alpha}{N}$
\end{Large} $[cm^2]$ is computed by the formula \eqref{eq:alphaN}
\begin{equation}
\frac{\alpha}{N} = 
\left\{
\begin{aligned}
	2.10^{-16}\cdot\exp\bigg( \frac{-7.248.10^{-15}}{||\vec{E}||/N}\bigg), \hspace{0.4cm} if \hspace{0.4cm} \frac{||\vec{E}||}{N} > 1.5\cdot10^{-15} \\ \\
	6.669^{-17}\cdot\exp\bigg( \frac{-5.593.10^{-15}}{||\vec{E}||/N}\bigg), \hspace{0.2cm} else \hspace{2.9cm}	
\end{aligned}
\right.
\label{eq:alphaN}
\end{equation}
\noindent With $N$, neutral gas density ($N$ = $2.5\cdot10^{19} cm^{-3}$). $C_1$ and $C_2$ are constants whose depend on \begin{Large}$\frac{||\vec{E}||}{N}$\end{Large} \cite{BENKHALDOUN20124623}.
\paragraph{Initial conditions}
The initial Gaussian pulse for the electron and the ion densities creates a disturbance in the electric field which is necessary for the initiation of the ionization wave propagation. The background electron and ion with a density of $10^{12}$ substitutes the photoionization phenomenon which is neglected in our simple discharge model.
\begin{equation}
\begin{aligned}
	n_e = 10^{16} \cdot \exp\left(-\frac{\left(x-0.2\right)^2 +  \left(y-0.25\right)^2 + \left(z-0.25\right)^2}{\sigma^2}\right)
	+ 10^{12} [cm^{-3}]^, \; \; \; \sigma = 0.01, \hspace{8.5cm}\\ \nonumber
	n_p = n_e	\hspace{19cm}	 \hspace{3cm}\nonumber
\end{aligned}
\end{equation}
\vspace{0.1cm}
\paragraph{Boundary conditions}
Dirichlet boundary conditions are applied for the potential $V$ ($V=12500[Volt]$ at the inlet boundary and $V=0[Volt]$ at the outlet), and homogeneous Neumann boundary conditions are applied otherwise.
Homogeneous Neumann boundary conditions are applied at all the boundaries for $n_e$, $n_p$, $\vec{V}$ and $\vec{E}$.
\paragraph{Data exchange}
For the 3D streamer model, the halo information sent each iteration are; electron density $n_e$, positive ion density $n_p$, potential $V$, electric field $\vec{E}(E_x, E_y, E_z)$, velocity field $\vec{v_e}({v_e}_x, {v_e}_y, {v_e}_z)$, cell gradient $\nabla n_e({n_e}_x, {n_e}_y, {n_e}_z)$ and $\psi$ coming from the barth jeperson method. The communications are performed using \textbf{MPI\_Neighbor\_alltoallv}, which allows sending data only to neighbor subdomains. Table \ref{table:neighbor} shows the computational time (s) for the communication part (one iteration for 3D Streamer Model) using \textbf{MPI\_Alltoallv}, \textbf{MPI\_Neighbor\_alltoallv} and \textbf{MPI\_Ineighbor\_alltoallv}. Table \ref{table:cpus} shows that the communication cost decreases with increasing number of MPI cores; the number of halo cells for each subdomain decreases (see table \ref{tab:metis_statistic}).
%
\begin{table}[h!]
\footnotesize
\renewcommand{\arraystretch}{1}
\setlength\tabcolsep{5pt}
\begin{center}
	\caption{Computation time (s) using different MPI functions.}\label{table:neighbor}
	\begin{tabular}{lccc}
		\hline
		\noalign{\smallskip}
		Cores  & Alltoallv & Neighbor\_alltoallv & Ineighbor\_alltoallv\\
		\noalign{\smallskip}
		\hline
		\noalign{\smallskip}
		128   & 6.74e-01 & 6.74e-01 & 6.76e-01 \\
		256   & 2.39e-01 & 2.20e-01 & 2.25e-01\\
		512   & 1.09e-01 & 7.39e-02 & 7.30e-02 \\
		1024  & 1.16e-01 & 2.31e-02 & 2.33e-02\\
		2048  & 2.71e-01 & 1.17e-02 & 1.23e-02\\
		4096  & 5.79e-01& 4.63e-03  & 4.89e-03\\
		8092  & 13.4e-01 & 3.08e-03 & 3.21e-03 \\
		16384 & 30.1e-01 & 2.49e-03 & 2.71e-03\\
		\hline
	\end{tabular}
\end{center}
\end{table}
\paragraph{Results and discussion}
Figure \ref{fig:iso2d} and \ref{fig:nobranch} and depict the isolines in a cut plane $y = 0.25$, and  3D paraview plot for electron density $n_e$ and net charge $n_e-n_p$.  The results aren't compared with laboratory experiments, but validated by our physics department. The computational time (s) for the different parts are presented in table \ref{table:cpus}, and shows that the most costly part is the Poisson equation's solving. A perfect scaling up to 8192 MPI cores is presented in figure \ref{fig:speedup_parts}, which validate our FV operators implementations, and the highly scalable PETSc solver. Unfortunately, in this test case, we did not take advantage of matrix assembly because its computed only once (the matrix coefficients depend only on the geometry).
\begin{figure}[h!]
\centering
\caption{Electron density (left) and Net charge density (right) at $t=20.5$ ns.}
\includegraphics[width=.8\textwidth]{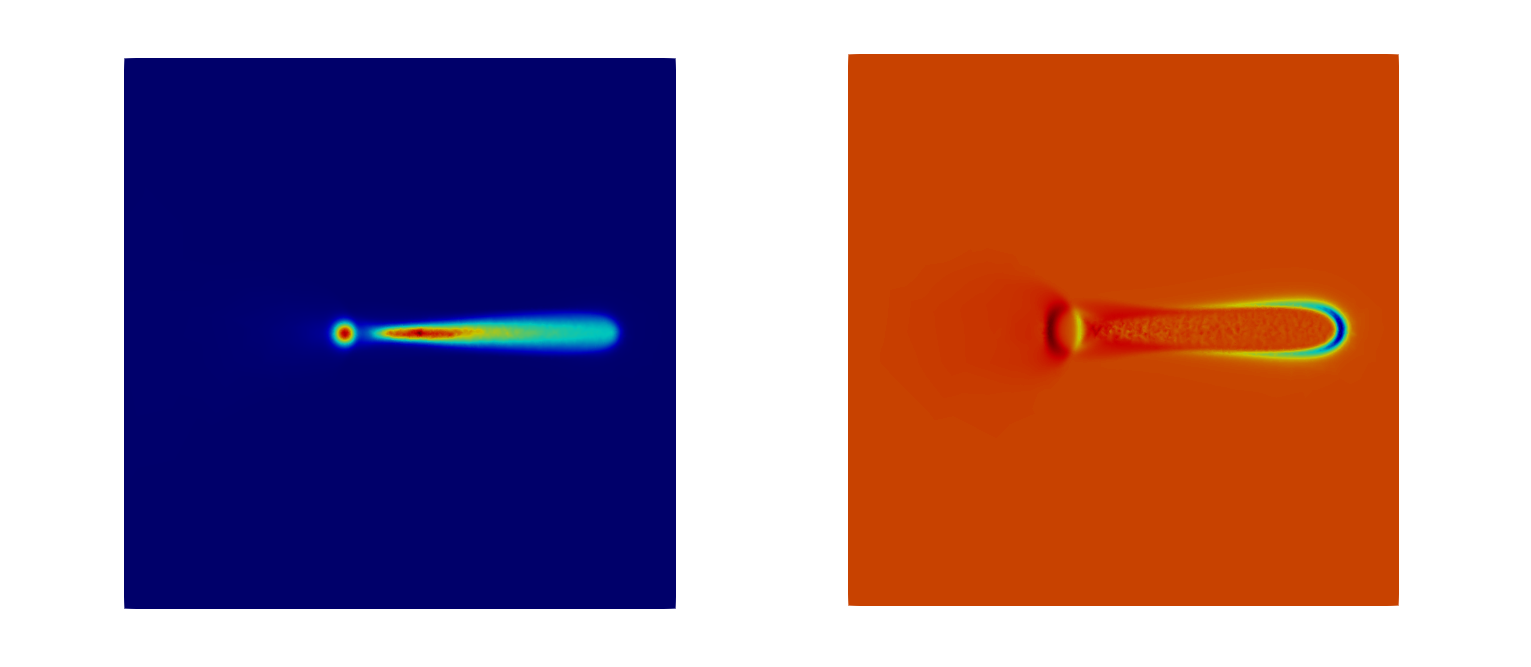}
\label{fig:iso2d}
\end{figure}
\begin{figure}[h!]
\centering
\caption{Electron density (left) and Net charge density (right) at $t=20.5$ ns.}
\includegraphics[width=.8\textwidth]{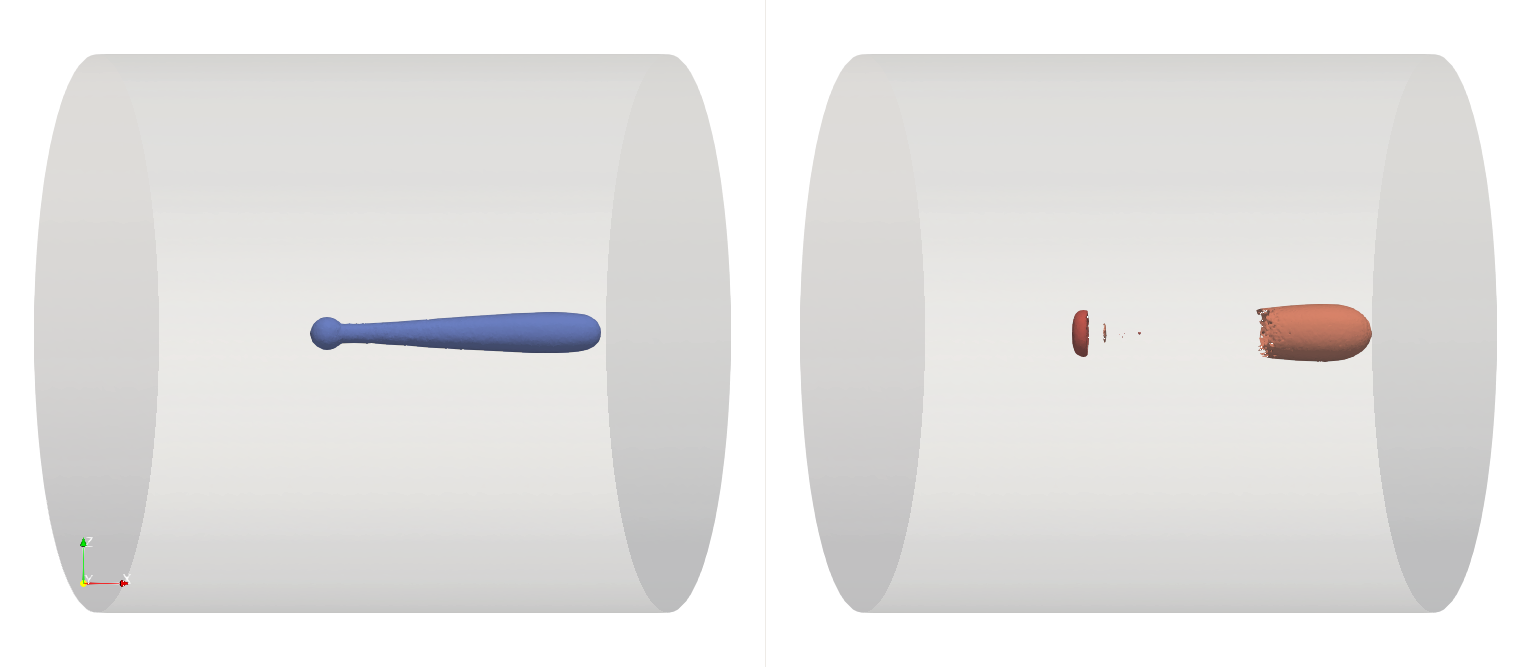}
\label{fig:nobranch}
\end{figure}

\begin{table*}[h!]
\footnotesize
\renewcommand{\arraystretch}{1}
\setlength\tabcolsep{6pt}
\begin{center}
	\caption{Computational time (s) for the different simulation's part.}\label{table:cpus}
	\begin{tabular}{lllllll}
		\hline
		\noalign{\smallskip}
		Cores & Cell Grad. & Face Grad. & Fluxes &  Least square &  PETSc (GAMG) & Communications \\
		\hline
		\noalign{\smallskip}
		1024 & 2957 & 322 & 1349 & 1144 & 1689798 & 578 \\
		2048 & 1158 & 131 & 589 & 538 & 804666 & 291 \\
		4096 & 507 & 66 & 282 & 199 & 371651 & 115 \\
		8192 & 234 & 47 & 148 & 79 & 218618 & 76 \\
	\end{tabular}
\end{center}
\end{table*}
\begin{figure}[h!]
\centering
\caption{Speedup for the different parts of the simulation.}
\includegraphics[width=0.7\textwidth]{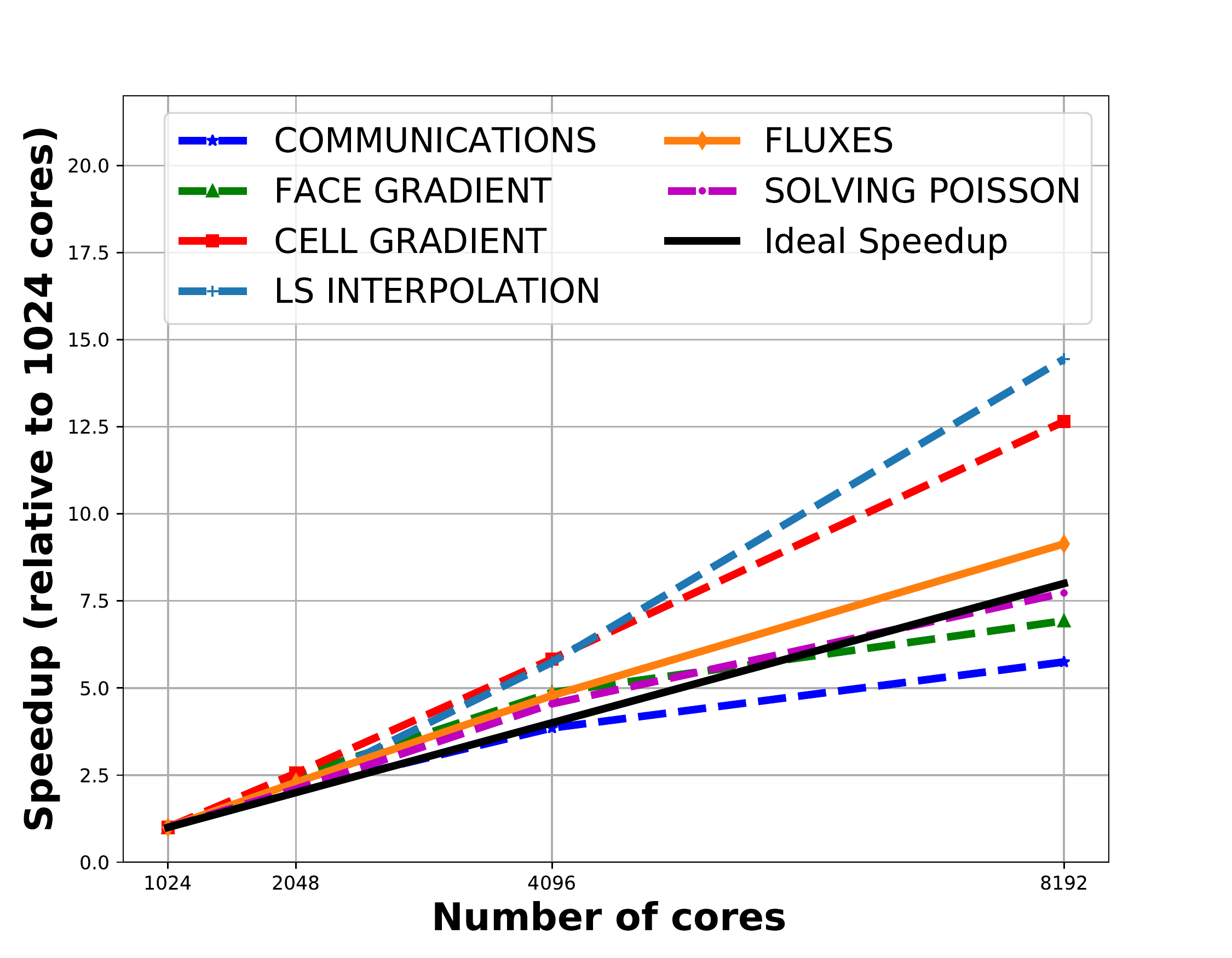}
\label{fig:speedup_parts}
\end{figure}

\subsubsection{Streamer Branching}
The plasma spot is added at time $t=1.26\cdot 10^{-8}$~s and canceled after the "duration time".

\noindent{\textit{$1^{st}$ plasma spot:}} $S_{e_t} = 10^{25}\cdot e^{\frac{\left(x-0.3\right)^2 + \left(y-0.25\right)^2 + \left(z-0.28\right)^2}{0.005^2}}$\\
{\textit{Position:}} $X_{01} = \left[0.3,\ 0.25,\ 0.28\right]$\\
{\textit{Duration:}} $t_1=0.5\cdot 10^{-9}$ s\\
{\textit{$2^{nd}$ plasma spot:}} $S_{e_t} = 10^{25}\cdot e^{\frac{\left(x-0.31\right)^2 + \left(y-0.25\right)^2 + \left(z-0.22\right)^2}{0.005^2}}$\\
{\textit{Position:}} $X_{02} = \left[0.31,\ 0.25,\ 0.22\right]$\\
{\textit{Duration:}} $t_2=0.5\cdot 10^{-9}$ s\\

\begin{figure}[h!]
	\centering
	\caption{Electron density (top) and Net charge density (bottom) using branching.}
	\includegraphics[width=0.495\textwidth]{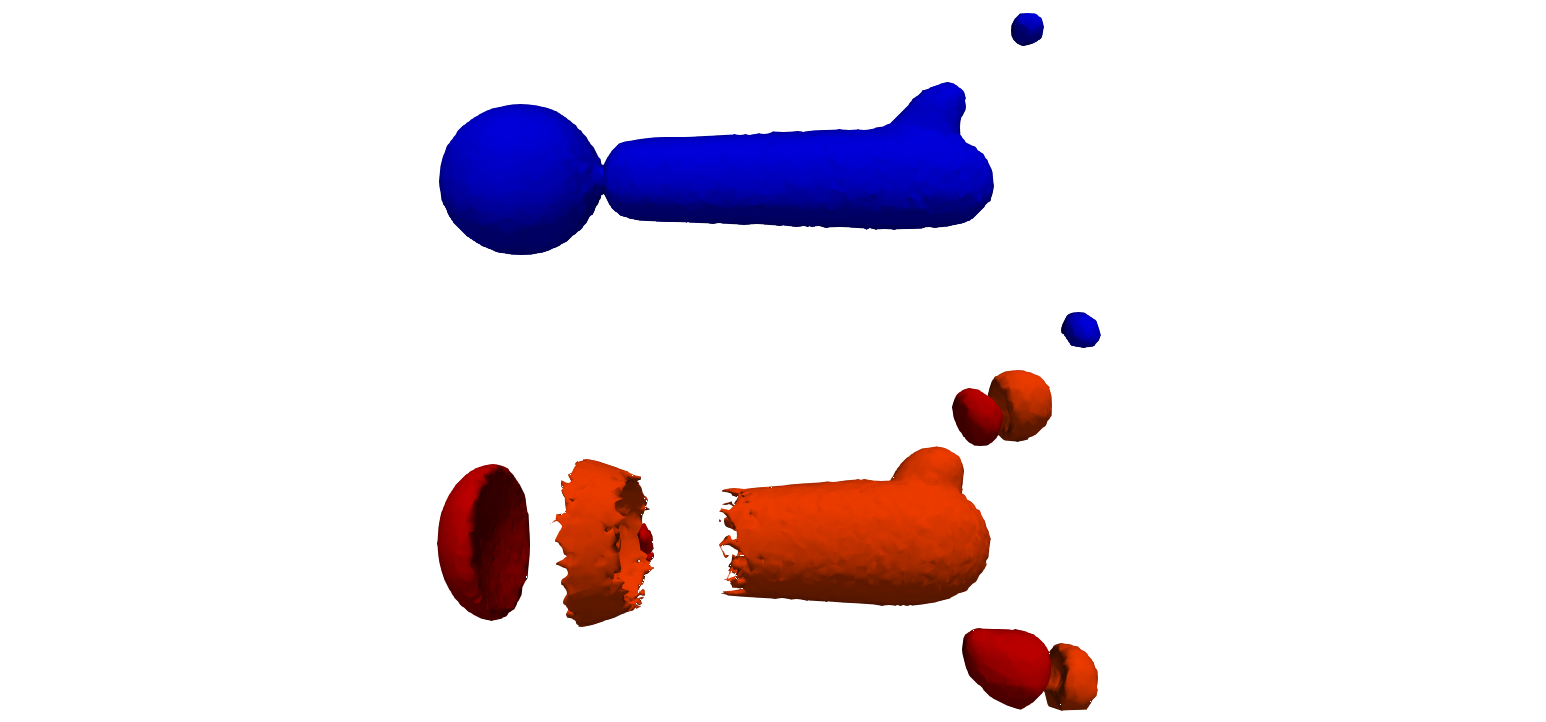}
	\includegraphics[width=0.495\textwidth]{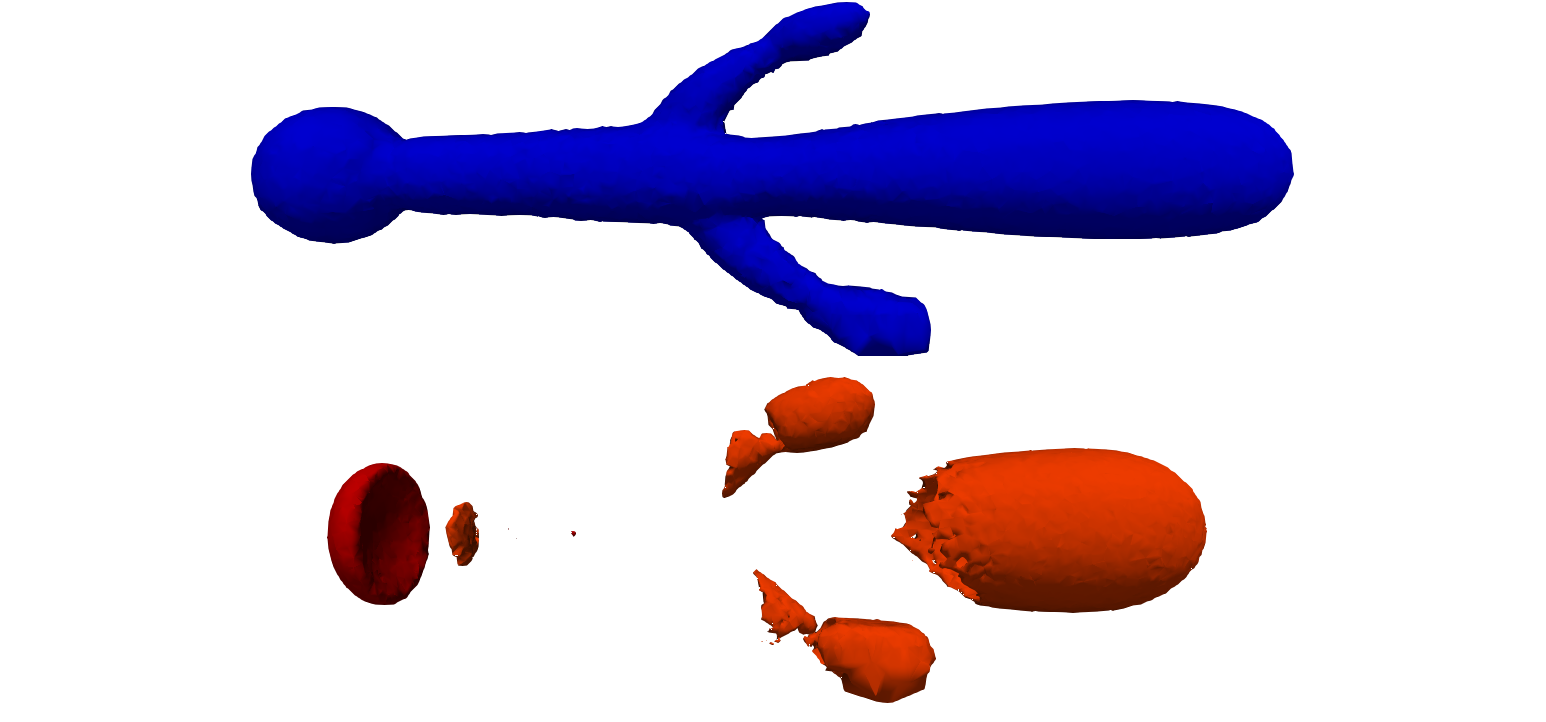}
	\label{fig:branchv4}
\end{figure}

\section{Conclusion}

In this work, we have presented the Manapy computational fluid dynamics (CFD) framework written using Python, giving the possibility to generate both C and Fortran subroutines using Pyccel, targeted at students, academics, and personal users to help them understand the general theory behind modern CFD solution methods and discretization technique, and also to deal with industrial problems. In addition, a new combination of physics can be implemented far faster than we ever could when we were coding in C or Fortran. 

Manapy addresses the need of many students and researchers to have a code easy to understand and to modify. A solver to use to test hypotheses but still able to deal with non-trivial geometries and complex flow physics.

Our current development efforts focus on allowing automatic code generation using an abstract form. Interesting efforts are being made to add the parallel adaptive mesh refinement (PAMR) procedure, improving performance, through parallelism and more sophisticated matrix preconditioners and solvers. We look forward to tackling even larger and more complicated problems, while keeping the ease of use that we've already established. 

\section{Annexes}

In this example, we solve equation \eqref{eq:manapyexmp} in 3D using gmsh file "cube.msh". We consider dirichlet boundary conditions on the inlet ($P=10.$), outlet ($P=0.$) and neumann boundary conditions otherwise.

\begin{equation}
\nabla^2 P = 0.
\label{eq:manapyexmp}
\end{equation}

\begin{itemize}
\item Imports 
\begin{python}
from manapy.ddm import readmesh
from manapy.ddm import Domain
from manapy.ast import Variable, LinearSystem
\end{python}
\item Read mesh and set up the local domains
\begin{python}
dim = 3
readmesh("cube.msh", dim=dim)

#Set up the domain
domain = Domain(dim=dim)
\end{python}

\item Add boundary conditions for variable $w$

\begin{python}
boundaries = {"in" : "dirichlet", "out" : "dirichlet",
	"upper":"neumann", "bottom":"neumann",
	"front":"neumann", "back":"neumann"}

values = {"in" : 10., "out": 0.}

P = Variable(domain=domain, BC=boundaries, values=values)
\end{python}

\item Initiate the linear system choosing MUMPS solver
\begin{python}
L = LinearSystem(domain=domain, var=P, solver="mumps")
\end{python}

\item Assembly the matrix
\begin{python}
L.assembly()
\end{python}

\item Solving the linear system
\begin{python}
L.solve(rhs=None) #or L.solve()
\end{python}


\item Saving result using paraview
\begin{python}
domain.save_on_cell(value=w.cell)
\end{python}

\item Compare with exact solution

\begin{python}
f = lambda x, y, z : 10. * (1. - x)
cells = domain.cells
nbcells = domain.nbcells

fexact = np.zeros(nbcells)
fexact[:] = f(cells.center[:][0], cells.center[!][1],  cells.center[:][2])

errorl2 = w.norml2(exact=fexact, order=1)  

print("l2 norm is ", errorl2)
\end{python}
\end{itemize}

\bibliographystyle{unsrt}
\bibliography{Draft}

\begin{thebibliography}{10}

\bibitem{chen2014openfoam}
Goong Chen, Qingang Xiong, Philip~J Morris, Eric~G Paterson, Alexey Sergeev,
  and Y~Wang.
\newblock Openfoam for computational fluid dynamics.
\newblock {\em Not. AMS}, 61(4):354--363, 2014.

\bibitem{solidworks2005solidworks}
Dassault~Syst{\`e}mes SolidWorks.
\newblock Solidworks{\textregistered}.
\newblock {\em Version Solidworks}, 2005.

\bibitem{fluent2015ansys}
ANSYS Fluent.
\newblock Ansys fluent.
\newblock {\em Academic Research. Release}, 14, 2015.

\bibitem{icenhour2018multi}
Casey Icenhour, Shane Keniley, Corey DeChant, Cody Permann, Alex Lindsay,
  Richard Martineau, Davide Curreli, and Steven Shannon.
\newblock Multi-physics object oriented simulation environment (moose).
\newblock Technical report, Idaho National Lab.(INL), Idaho Falls, ID (United
  States), 2018.

\bibitem{permann2020moose}
Cody~J. Permann, Derek~R. Gaston, David Andr{\v{s}}, Robert~W. Carlsen, Fande
  Kong, Alexander~D. Lindsay, Jason~M. Miller, John~W. Peterson, Andrew~E.
  Slaughter, Roy~H. Stogner, and Richard~C. Martineau.
\newblock {MOOSE}: Enabling massively parallel multiphysics simulation.
\newblock {\em {SoftwareX}}, 11:100430, 2020.

\bibitem{alnaes2015fenics}
Martin Aln{\ae}s, Jan Blechta, Johan Hake, August Johansson, Benjamin Kehlet,
  Anders Logg, Chris Richardson, Johannes Ring, Marie~E Rognes, and Garth~N
  Wells.
\newblock The fenics project version 1.5.
\newblock {\em Archive of Numerical Software}, 3(100), 2015.

\bibitem{richardson_wells_2015}
Chris N.~Richardson and Garth~N. Wells.
\newblock Parallel scaling of dolfin on archer, Feb 2015.

\bibitem{CimrmanLukesRohan}
Robert Cimrman, Vladimír Lukeš, and Eduard Rohan.
\newblock Multiscale finite element calculations in python using sfepy.
\newblock {\em Advances in Computational Mathematics}, 2019.

\bibitem{Lukes2020HomogenizationOL}
Vladim{\'i}r Lukes and Eduard Rohan.
\newblock Homogenization of large deforming fluid-saturated porous structures.
\newblock {\em ArXiv}, abs/2012.03730, 2020.

\bibitem{Rohan2021HomogenizationOT}
Eduard Rohan and Vladim{\'i}r Lukes.
\newblock Homogenization of the vibro-acoustic transmission on periodically
  perforated elastic plates with arrays of resonators.
\newblock {\em ArXiv}, abs/2104.01367, 2021.

\bibitem{BARTUSCHAT2018147}
Dominik Bartuschat and Ulrich Rüde.
\newblock A scalable multiphysics algorithm for massively parallel direct
  numerical simulations of electrophoretic motion.
\newblock {\em Journal of Computational Science}, 27:147--167, 2018.

\bibitem{BAUER2021478}
Martin Bauer, Sebastian Eibl, Christian Godenschwager, Nils Kohl, Michael
  Kuron, Christoph Rettinger, Florian Schornbaum, Christoph Schwarzmeier,
  Dominik Thönnes, Harald Köstler, and Ulrich Rüde.
\newblock walberla: A block-structured high-performance framework for
  multiphysics simulations.
\newblock {\em Computers and Mathematics with Applications}, 81:478--501, 2021.
\newblock Development and Application of Open-source Software for Problems with
  Numerical PDEs.

\bibitem{Fipy}
Jonathan~E. Guyer, Daniel Wheeler, and James~A. Warren.
\newblock Fipy: Partial differential equations with python.
\newblock {\em Computing in Science Engineering}, 11(3):6--15, 2009.

\bibitem{ALINOVI2021100655}
Edoardo Alinovi and Joel Guerrero.
\newblock Flubio—an unstructured, parallel, finite-volume based
  navier–stokes and convection–diffusion like equations solver for teaching
  and research purposes.
\newblock {\em SoftwareX}, 13:100655, 2021.

\bibitem{10.1145/2833157.2833162}
Siu~Kwan Lam, Antoine Pitrou, and Stanley Seibert.
\newblock Numba: A llvm-based python jit compiler.
\newblock In {\em Proceedings of the Second Workshop on the LLVM Compiler
  Infrastructure in HPC}, LLVM '15, New York, NY, USA, 2015. Association for
  Computing Machinery.

\bibitem{Karypis:1998:FHQ:305219.305248}
George Karypis and Vipin Kumar.
\newblock A fast and high quality multilevel scheme for partitioning irregular
  graphs.
\newblock {\em SIAM J. Sci. Comput.}, 20(1):359--392, December 1998.

\bibitem{DALCIN20111124}
Lisandro~D. Dalcin, Rodrigo~R. Paz, Pablo~A. Kler, and Alejandro Cosimo.
\newblock Parallel distributed computing using python.
\newblock {\em Advances in Water Resources}, 34(9):1124--1139, 2011.
\newblock New Computational Methods and Software Tools.

\bibitem{osti_1483828}
S.~Balay, S.~Abhyankar, M.~Adams, J.~Brown, P.~Brune, K.~Buschelman, L.~Dalcin,
  A.~Dener, V.~Eijkhout, W.~Gropp, D.~Karpeyev, D.~Kaushik, M.~Knepley, D.~May,
  L.~Curfman McInnes, R.~Mills, T.~Munson, K.~Rupp, P.~Sanan, B.~Smith,
  S.~Zampini, H.~Zhang, and H.~Zhang.
\newblock Petsc users manual: Revision 3.10.
\newblock 9 2018.

\bibitem{4160265}
John~D. Hunter.
\newblock Matplotlib: A 2d graphics environment.
\newblock {\em Computing in Science Engineering}, 9(3):90--95, 2007.

\bibitem{AHRENS2005717}
JAMES AHRENS, BERK GEVECI, and CHARLES LAW.
\newblock 36 - paraview: An end-user tool for large-data visualization.
\newblock In Charles~D. Hansen and Chris~R. Johnson, editors, {\em
  Visualization Handbook}, pages 717--731. Butterworth-Heinemann, Burlington,
  2005.

\bibitem{10.1007/978-3-030-43651-3_42}
Moussa Ziggaf, Mohamed Boubekeur, Imad kissami, Fayssal Benkhaldoun, and
  Imad~El Mahi.
\newblock The fvc scheme on unstructured meshes for the two-dimensional shallow
  water equations.
\newblock In Robert Kl{\"o}fkorn, Eirik Keilegavlen, Florin~A. Radu, and
  J{\"u}rgen Fuhrmann, editors, {\em Finite Volumes for Complex Applications IX
  - Methods, Theoretical Aspects, Examples}, pages 455--465, Cham, 2020.
  Springer International Publishing.

\bibitem{atmos11040314}
Arakel Petrosyan, Dmitry Klimachkov, Maria Fedotova, and Timofey Zinyakov.
\newblock Shallow water magnetohydrodynamics in plasma astrophysics. waves,
  turbulence, and zonal flows.
\newblock {\em Atmosphere}, 11(4), 2020.

\bibitem{BENKHALDOUN20124623}
Fayssal Benkhaldoun, Jaroslav Fořt, Khaled Hassouni, and Jan Karel.
\newblock Simulation of planar ionization wave front propagation on an
  unstructured adaptive grid.
\newblock {\em Journal of Computational and Applied Mathematics},
  236(18):4623--4634, 2012.
\newblock FEMTEC 2011: 3rd International Conference on Computational Methods in
  Engineering and Science, May 9–13, 2011.

\bibitem{Fort2019}
J.~Fo{\v{r}}t, J.~Karel, D.~Trdli{\v{c}}ka, F.~Benkhaldoun, I.~Kissami, J.-B.
  Montavon, K.~Hassouni, and J.~Zs. Mezei.
\newblock Finite volume methods for numerical simulation of the discharge
  motion described by different physical models.
\newblock {\em Advances in Computational Mathematics}, 45(4):2163--2189, Aug
  2019.

\bibitem{ISSA198640}
R.I Issa.
\newblock Solution of the implicitly discretised fluid flow equations by
  operator-splitting.
\newblock {\em Journal of Computational Physics}, 62(1):40--65, 1986.

\bibitem{_ern_k_2011}
M~{\v{C}}ern{\'{a}}k, D~Kov{\'{a}}{\v{c}}ik, J~R{\'{a}}hel{\textquotesingle},
  P~St{\textquotesingle}ahel, A~Zahoranov{\'{a}}, J~Kubincov{\'{a}},
  A~T{\'{o}}th, and L{\textquotesingle}~{\v{C}}ern{\'{a}}kov{\'{a}}.
\newblock Generation of a high-density highly non-equilibrium air plasma for
  high-speed large-area flat surface processing.
\newblock {\em Plasma Physics and Controlled Fusion}, 53(12):124031, nov 2011.

\bibitem{Joshi2013}
Ravindra~P. Joshi and Selma~Mededovic Thagard.
\newblock Streamer-like electrical discharges in water: Part ii. environmental
  applications.
\newblock {\em Plasma Chemistry and Plasma Processing}, 33(1):17--49, Feb 2013.

\bibitem{NOZAKI201329}
Tomohiro Nozaki and Ken Okazaki.
\newblock Non-thermal plasma catalysis of methane: Principles, energy
  efficiency, and applications.
\newblock {\em Catalysis Today}, 211:29--38, 2013.
\newblock Recent Advances in Plasma and Catalysis (ISPCEM 2012).

\bibitem{879360}
H.~Akiyama.
\newblock Streamer discharges in liquids and their applications.
\newblock {\em IEEE Transactions on Dielectrics and Electrical Insulation},
  7(5):646--653, 2000.

\end{thebibliography}

\end{document}